\PassOptionsToPackage{table}{xcolor}
\documentclass[conference]{IEEEtran}

\usepackage{booktabs}
\usepackage{balance}
\let\labelindent\relax
\usepackage[inline]{enumitem}
\usepackage[hyphens]{url}
\usepackage{hyperref}
\usepackage[hyphenbreaks]{breakurl}
\usepackage{multirow}
\usepackage{url}
\usepackage{pgf-pie}
\usepackage{pgfplots} 
\pgfplotsset{width=6.5cm, compat=1.6}  
\usepackage{tikz}
\usepackage{xcolor}
\hypersetup{%
  pdflang={en-US},
  pdftitle={A Transcontinental Analysis of Account Remediation Protocols of Popular Websites},
  pdfauthor={Philipp Markert, Andrick Adhikari, Sanchari Das},
  pdfsubject={},
  pdfkeywords={account remediation},
  pdfdisplaydoctitle=true, 
  pdfstartview={FitH},
  breaklinks
}

\definecolor{bblue}{HTML}{4F81BD}
\definecolor{rred}{HTML}{C0504D}
\definecolor{ggreen}{HTML}{9BBB59}
\definecolor{ppurple}{HTML}{9F4C7C}
\definecolor{Mycolor2}{HTML}{00F9DE}

\begin{document}

\title{A Transcontinental Analysis of Account Remediation Protocols of Popular Websites}

\author{\IEEEauthorblockN{Philipp Markert}
        \IEEEauthorblockA{Ruhr University Bochum\\
        philipp.markert@rub.de}
\and
        \IEEEauthorblockN{Andrick Adhikari}
        \IEEEauthorblockA{University of Denver\\
        andrick.adhikari@du.edu}
\and
        \IEEEauthorblockN{Sanchari Das}
        \IEEEauthorblockA{University of Denver\\
        sanchari.das@du.edu}
}

\IEEEoverridecommandlockouts
\makeatletter\def\@IEEEpubidpullup{6.5\baselineskip}\makeatother
\IEEEpubid{\parbox{\columnwidth}{
    Symposium on Usable Security and Privacy (USEC) 2023 \\
    27 February 2023, San Diego, CA, USA \\
    ISBN 1-891562-91-6 \\
    https://dx.doi.org/10.14722/usec.2023.235078 \\
    www.ndss-symposium.org, https://www.usablesecurity.net/USEC/
}
\hspace{\columnsep}\makebox[\columnwidth]{}}

\maketitle

\newcommand{\tableCountries}[0]{
\begin{table}[tb!]
\caption{List of the analyzed countries for each of the 5 continents. The Global Cybersecurity Index~\cite{index2020url} was used for the classification of best and worst countries.} 
\centering
\label{tab:countries}
\begin{tabular}{ l | l | l } 
\toprule
\textbf{Continent} & \textbf{Best Countries} & \textbf{Worst Countries} \\
\midrule
& Mauritius & Equatorial Guinea \\ 
Africa & Tanzania & Eritrea \\ 
&  Ghana & Burundi\\ 
\midrule
& USA & Honduras \\ 
America & Canada & Dominica \\ 
&  Brazil &  Haiti\\ 
\midrule
& Saudi Arabia & Maldives \\ 
Asia & South Korea & Timor-Leste \\ 
&  Singapore & Afghanistan\\ 
\midrule
& United Kingdom & San Marino \\ 
Europe & Estonia & Andorra\\ 
&  Spain & Bosnia and Herzegovina\\ 
\midrule
& Australia & Marshall Islands \\ 
Oceania & New Zealand & Solomon Islands \\ 
&  Fiji &  Vanuatu\\ 
\bottomrule
\end{tabular}
\end{table}
}

\newcommand{\tableAnalysisCombined}[0]{
\begin{table*}[ht!]
\renewcommand{\arraystretch}{1.1}
\centering
   \caption{Percentage of websites that provide advice for the 5 phases of account remediation among the top 50 websites of the 3 best and worst ranked countries according to the GCI~\cite{index2020url} in Africa, America, Asia, Europe, and Oceania.}\label{tab:analysis}
   \resizebox{\textwidth}{!}{\begin{tabular}{@{}>{\cellcolor{white}} cr | p{2.25em}<{\centering}p{2.25em}<{\centering} p{2.25em}<{\centering}p{2.25em}<{\centering} p{2.25em}<{\centering}p{2.25em}<{\centering} p{2.25em}<{\centering}p{2.25em}<{\centering} p{2.25em}<{\centering}p{2.25em}<{\centering}}
   \toprule
   & & \multicolumn{2}{c}{\textbf{Africa}} & \multicolumn{2}{c}{\textbf{America}} & \multicolumn{2}{c}{\textbf{Asia}} & \multicolumn{2}{c}{\textbf{Europe}} & \multicolumn{2}{c}{\textbf{Oceania}} \\
   & & Best & Worst & Best & Worst & Best & Worst & Best & Worst & Best & Worst \\
   \midrule
   & Account Locked by Provider & 50\% & 58\% & 55\% & 57\% & 58\% & 57\% & 58\% & 57\% & 49\% & 63\% \\
   \rowcolor{lightgray!50} & Account Otherwise Unavailable & 51\% & 57\% & 58\% & 52\% & 60\% & 52\% & 60\% & 53\% & 49\% & 47\% \\
   & Billing/Finance Issues & 57\% & 43\% & 63\% & 58\% & 63\% & 54\% & 64\% & 58\% & 59\% & 32\% \\
   \rowcolor{lightgray!50} & Email Changed & 37\% & 35\% & 39\% & 39\% & 41\% & 37\% & 38\% & 41\% & 37\% & 38\% \\
   & Explicit Notification & 83\% & 80\% & 82\% & 91\% & 82\% & 80\% & 82\% & 89\% & 78\% & 73\% \\
   \rowcolor{lightgray!50} & Observed Unauthorized Logins & 49\% & 42\% & 55\% & 54\% & 58\% & 52\% & 58\% & 51\% & 54\% & 39\% \\
   & Password Changed & 42\% & 56\% & 43\% & 47\% & 43\% & 48\% & 42\% & 48\% & 46\% & 63\% \\
   \rowcolor{lightgray!50} & Social Media or Third-Party Account Connected & 22\% & 25\% & 23\% & 19\% & 23\% & 19\% & 22\% & 18\% & 25\% & 38\% \\
   \multirow{-9}{*}{\rotatebox{90}{\scriptsize{\textbf{Compromise Discovery}}}} & Unauthorized/Suspicious Activity & 48\% & 48\% & 55\% & 51\% & 58\% & 50\% & 58\% & 50\% & 51\% & 47\% \\
   \midrule
   & Customer Service Process & 83\% & 78\% & 82\% & 78\% & 85\% & 78\% & 84\% & 80\% & 79\% & 84\% \\
   \rowcolor{lightgray!50} & Password Reset & 88\% & 92\% & 88\% & 92\% & 88\% & 88\% & 90\% & 92\% & 89\% & 92\% \\
   \multirow{-3.15}{*}{\rotatebox{90}{\scriptsize{\textbf{\shortstack{Account\\Recovery}}}}} & Password Rotation & 45\% & 38\% & 55\% & 45\% & 58\% & 44\% & 58\% & 46\% & 46\% & 38\% \\
   \midrule
   & Instructions for Account Deletion & 79\% & 70\% & 87\% & 80\% & 87\% & 78\% & 87\% & 78\% & 80\% & 71\% \\
   \rowcolor{lightgray!50} & Remove Third Party Access & 55\% & 48\% & 58\% & 51\% & 60\% & 50\% & 59\% & 50\% & 57\% & 47\% \\
   & Review Active Session & 42\% & 38\% & 45\% & 43\% & 48\% & 43\% & 48\% & 42\% & 43\% & 40\% \\
   \rowcolor{lightgray!50} & Sign Out Everywhere (Specific Function) & 28\% & 19\% & 28\% & 29\% & 27\% & 26\% & 27\% & 28\% & 29\% & 26\% \\
   \multirow{-5}{*}{\rotatebox{90}{\scriptsize{\textbf{\shortstack{Limit \\Unwanted \\Access}}}}} & Sign Out of Unknown Sessions & 28\% & 35\% & 35\% & 29\% & 36\% & 31\% & 36\% & 29\% & 31\% & 43\% \\
   \midrule
   & Customer Service Process & 46\% & 50\% & 48\% & 44\% & 48\% & 45\% & 47\% & 45\% & 46\% & 60\% \\
   \rowcolor{lightgray!50} & Fix Logs of Past Viewing/Activity/Content History & 18\% & 18\% & 20\% & 19\% & 20\% & 19\% & 19\% & 18\% & 21\% & 22\% \\
   & Review and/or Remove Activities/Content & 26\% & 27\% & 28\% & 25\% & 28\% & 22\% & 27\% & 23\% & 28\% & 34\% \\
   \rowcolor{lightgray!50} & Verify Settings & 64\% & 65\% & 66\% & 59\% & 69\% & 57\% & 68\% & 57\% & 60\% & 71\% \\
   \multirow{-5}{*}{\rotatebox{90}{\scriptsize{\textbf{\shortstack{Service\\Restoration}}}}} & Verify User Information & 60\% & 58\% & 63\% & 61\% & 66\% & 57\% & 65\% & 60\% & 59\% & 74\% \\
   \midrule
   & Advice About Secure Email & 58\% & 62\% & 58\% & 60\% & 61\% & 58\% & 61\% & 58\% & 57\% & 56\% \\
   \rowcolor{lightgray!50} & Always Log Out on Shared Devices & 27\% & 38\% & 33\% & 29\% & 33\% & 31\% & 33\% & 30\% & 30\% & 44\% \\
   & Check/Modify Related Accounts & 25\% & 17\% & 25\% & 24\% & 25\% & 23\% & 25\% & 24\% & 22\% & 26\% \\
   \rowcolor{lightgray!50} & Enable 2FA & 63\% & 65\% & 73\% & 65\% & 76\% & 63\% & 75\% & 67\% & 67\% & 62\% \\
   & Keep Software Up To Date & 36\% & 32\% & 38\% & 36\% & 41\% & 35\% & 41\% & 36\% & 34\% & 32\% \\
   \rowcolor{lightgray!50} & Password Advice: Strong, Unique, Change & 75\% & 56\% & 76\% & 79\% & 79\% & 73\% & 78\% & 74\% & 75\% & 47\% \\
   & Regular Security Checkups/Advice & \hspace{.5em}5\% & \hspace{.5em}5\% & \hspace{.5em}8\% & \hspace{.5em}5\% & \hspace{.5em}8\% & \hspace{.5em}5\% & \hspace{.5em}8\% & \hspace{.5em}6\% & \hspace{.5em}4\% & 10\% \\
   \rowcolor{lightgray!50} & Remove Access to Third-Party Apps & 34\% & 22\% & 35\% & 33\% & 35\% & 32\% & 34\% & 31\% & 35\% & 19\% \\
   \multirow{-9}{*}{\rotatebox{90}{\scriptsize{\textbf{Prevention}}}} & Sign Out of Devices & 30\% & 24\% & 33\% & 29\% & 32\% & 31\% & 32\% & 28\% & 30\% & 36\% \\
   \bottomrule
   \end{tabular}}
\end{table*}}

\newcommand{\tableAllWebsitesPartOne}[0]{
\begin{table*}[h!]
\centering
\scriptsize
\caption{The number of times each of the 158 websites appears in the top 50 of the 3 best and worst ranking nations, as determined by the GCI~\cite{index2020url} in each of the continents. Websites with the same total were ordered alphabetically.} 
\label{tab:allwebsitesPartOne}
\rowcolors{3}{white}{lightgray!50}
\resizebox{.74\columnwidth}{!}{
\begin{tabular}{r p{9em}<{\raggedleft} | p{2.25em}<{\centering}p{2.25em}<{\centering} p{2.25em}<{\centering}p{2.25em}<{\centering} p{2.25em}<{\centering}p{2.25em}<{\centering} p{2.25em}<{\centering}p{2.25em}<{\centering} p{2.25em}<{\centering}p{2.25em}<{\centering} | c }
   \toprule
    && \multicolumn{2}{c}{\textbf{Africa}} & \multicolumn{2}{c}{\textbf{America}} & \multicolumn{2}{c}{\textbf{Asia}} & \multicolumn{2}{c}{\textbf{Europe}} & \multicolumn{2}{c|}{\textbf{Oceania}} \\
    \textbf{No.} & \multicolumn{1}{c|}{\textbf{Website}} & Best & Worst & Best & Worst & Best & Worst & Best & Worst & Best & Worst & \textbf{Total}\\
\midrule
1 & amazon.com & 3 & 2 & 3 & 3 & 3 & 3 & 3 & 3 & 3 & 3 & 29\\
2 & google.com & 2 & 3 & 3 & 3 & 3 & 3 & 3 & 3 & 3 & 3 & 29\\
3 & youtube.com & 2 & 3 & 3 & 3 & 3 & 3 & 3 & 3 & 3 & 3 & 29\\

4 & facebook.com & 2 & 2 & 3 & 3 & 3 & 3 & 3 & 3 & 3 & 3 & 28\\
5 & imdb.com & 3 & 2 & 3 & 3 & 2 & 3 & 3 & 3 & 3 & 3 & 28\\
6 & pinterest.com & 3 & 2 & 3 & 3 & 3 & 3 & 3 & 3 & 3 & 2 & 28\\
7 & twitter.com & 2 & 2 & 3 & 3 & 3 & 3 & 3 & 3 & 3 & 3 & 28\\
8 & wikipedia.org & 2 & 2 & 3 & 3 & 3 & 3 & 3 & 3 & 3 & 3 & 28\\

9 & linkedin.com & 2 & 2 & 3 & 3 & 3 & 3 & 3 & 3 & 3 & 2 & 27\\
10 & yahoo.com & 3 & 2 & 3 & 3 & 3 & 3 & 3 & 3 & 3 & 1 & 27\\

11 & instagram.com & 2 & 2 & 3 & 3 & 3 & 3 & 3 & 3 & 3 & 1 & 26\\
12 & live.com & 3 & 2 & 3 & 3 & 3 & 2 & 3 & 3 & 3 & 1 & 26\\
13 & microsoftonline.com & 3 & 2 & 3 & 3 & 3 & 3 & 3 & 3 & 1 & 2 & 26\\
14 & reddit.com & 3 & 1 & 3 & 3 & 3 & 3 & 3 & 3 & 3 & 1 & 26\\
15 & whatsapp.com & 3 & 2 & 3 & 3 & 3 & 3 & 3 & 3 & 3 & 0 & 26\\

16 & github.com & 3 & 1 & 3 & 3 & 3 & 2 & 3 & 3 & 3 & 1 & 25\\
17 & microsoft.com & 2 & 2 & 3 & 3 & 3 & 3 & 3 & 3 & 3 & 0 & 25\\
18 & netflix.com & 2 & 2 & 3 & 3 & 3 & 3 & 3 & 3 & 1 & 2 & 25\\

19 & bing.com & 3 & 2 & 3 & 3 & 3 & 2 & 3 & 2 & 3 & 0 & 24\\
20 & zoom.us & 3 & 1 & 3 & 3 & 3 & 2 & 3 & 3 & 2 & 1 & 24\\

21 & adobe.com & 3 & 0 & 3 & 3 & 3 & 2 & 3 & 3 & 3 & 0 & 23\\
22 & nih.gov & 3 & 1 & 3 & 3 & 3 & 2 & 3 & 2 & 3 & 0 & 23\\
23 & nytimes.com & 3 & 1 & 3 & 3 & 3 & 2 & 3 & 2 & 3 & 0 & 23\\
24 & paypal.com & 2 & 1 & 3 & 3 & 3 & 2 & 3 & 3 & 3 & 0 & 23\\
25 & spotify.com & 3 & 0 & 3 & 3 & 3 & 2 & 3 & 3 & 3 & 0 & 23\\

26 & apple.com & 2 & 0 & 3 & 3 & 3 & 2 & 3 & 3 & 3 & 0 & 22\\
27 & office.com & 3 & 0 & 3 & 3 & 3 & 2 & 3 & 2 & 3 & 0 & 22\\
28 & vimeo.com & 3 & 0 & 3 & 3 & 3 & 1 & 3 & 3 & 3 & 0 & 22\\
29 & vk.com & 3 & 1 & 3 & 2 & 3 & 2 & 3 & 2 & 3 & 0 & 22\\

30 & blogspot.com & 3 & 1 & 3 & 2 & 3 & 2 & 3 & 2 & 2 & 0 & 21\\
31 & msn.com & 3 & 1 & 0 & 3 & 3 & 2 & 3 & 3 & 3 & 0 & 21\\

32 & medium.com & 3 & 0 & 3 & 3 & 2 & 2 & 2 & 2 & 3 & 0 & 20\\
33 & tiktok.com & 0 & 2 & 3 & 3 & 3 & 2 & 3 & 3 & 0 & 1 & 20\\

34 & stackoverflow.com & 2 & 1 & 3 & 2 & 2 & 2 & 3 & 3 & 1 & 0 & 19\\

35 & mozilla.org & 3 & 0 & 3 & 2 & 3 & 1 & 3 & 2 & 1 & 0 & 18\\
36 & tumblr.com & 3 & 0 & 3 & 1 & 3 & 2 & 3 & 1 & 2 & 0 & 18\\

37 & soundcloud.com & 3 & 2 & 0 & 3 & 0 & 2 & 0 & 1 & 3 & 2 & 16\\

38 & ebay.com & 3 & 1 & 0 & 3 & 0 & 2 & 0 & 2 & 3 & 1 & 15\\
39 & sohu.com & 3 & 0 & 3 & 0 & 3 & 0 & 3 & 0 & 2 & 1 & 15\\

40 & aliexpress.com & 3 & 2 & 0 & 3 & 0 & 2 & 0 & 3 & 1 & 0 & 14\\
41 & baidu.com & 2 & 0 & 3 & 0 & 3 & 0 & 3 & 1 & 1 & 1 & 14\\
42 & qq.com & 1 & 0 & 3 & 0 & 3 & 0 & 3 & 1 & 2 & 1 & 14\\
43 & taobao.com & 1 & 0 & 3 & 0 & 3 & 0 & 3 & 1 & 2 & 1 & 14\\
44 & yandex.ru & 0 & 0 & 3 & 1 & 3 & 2 & 3 & 2 & 0 & 0 & 14\\

45 & cnn.com & 3 & 0 & 0 & 3 & 0 & 2 & 0 & 2 & 3 & 0 & 13\\
46 & flickr.com & 3 & 0 & 2 & 1 & 3 & 1 & 1 & 2 & 0 & 0 & 13\\
47 & theguardian.com & 2 & 1 & 0 & 2 & 0 & 3 & 0 & 1 & 3 & 1 & 13\\
48 & twitch.tv & 3 & 0 & 0 & 3 & 0 & 2 & 0 & 3 & 2 & 0 & 13\\
49 & weibo.com & 1 & 0 & 3 & 0 & 3 & 0 & 3 & 0 & 2 & 1 & 13\\

50 & dropbox.com & 3 & 0 & 0 & 2 & 0 & 2 & 0 & 2 & 3 & 0 & 12\\
51 & wordpress.org & 0 & 0 & 3 & 1 & 3 & 1 & 3 & 1 & 0 & 0 & 12\\

52 & amazonaws.com & 0 & 0 & 3 & 1 & 3 & 0 & 2 & 0 & 2 & 0 & 11\\
53 & bilibili.com & 0 & 0 & 3 & 0 & 3 & 0 & 3 & 1 & 0 & 1 & 11\\
54 & csdn.net & 1 & 0 & 3 & 0 & 3 & 0 & 3 & 0 & 0 & 1 & 11\\
55 & sina.com.cn & 1 & 0 & 3 & 0 & 2 & 0 & 3 & 0 & 2 & 0 & 11\\

56 & zhihu.com & 0 & 0 & 3 & 0 & 3 & 0 & 3 & 0 & 0 & 1 & 10\\

57 & canva.com & 0 & 1 & 0 & 3 & 0 & 2 & 0 & 3 & 0 & 0 & \hspace{.5em}9\\
58 & jd.com & 0 & 0 & 3 & 0 & 3 & 0 & 3 & 0 & 0 & 0 & \hspace{.5em}9\\

59 & etsy.com & 1 & 0 & 0 & 2 & 0 & 0 & 0 & 2 & 3 & 0 & \hspace{.5em}8\\
60 & europa.eu & 2 & 0 & 0 & 1 & 2 & 1 & 1 & 1 & 0 & 0 & \hspace{.5em}8\\
61 & forbes.com & 2 & 1 & 0 & 2 & 0 & 1 & 0 & 1 & 1 & 0 & \hspace{.5em}8\\
52 & quora.com & 0 & 1 & 0 & 2 & 0 & 2 & 0 & 1 & 0 & 2 & \hspace{.5em}8\\

63 & alibaba.com & 0 & 2 & 0 & 0 & 0 & 1 & 0 & 1 & 0 & 3 & \hspace{.5em}7\\

64 & bit.ly & 0 & 0 & 3 & 0 & 1 & 0 & 2 & 0 & 0 & 0 & \hspace{.5em}6\\
65 & healthline.com & 0 & 2 & 0 & 0 & 0 & 1 & 0 & 0 & 0 & 3 & \hspace{.5em}6\\
66 & icloud.com & 0 & 0 & 0 & 2 & 0 & 2 & 0 & 2 & 0 & 0 & \hspace{.5em}6\\
67 & wordpress.com & 3 & 0 & 0 & 0 & 0 & 1 & 0 & 0 & 2 & 0 & \hspace{.5em}6\\

68 & archive.org & 1 & 0 & 0 & 0 & 0 & 2 & 0 & 0 & 2 & 0 & \hspace{.5em}5\\
69 & imgur.com & 1 & 0 & 0 & 0 & 0 & 0 & 0 & 0 & 3 & 0 & \hspace{.5em}5\\
70 & mediafire.com & 0 & 2 & 0 & 0 & 0 & 1 & 0 & 0 & 0 & 2 & \hspace{.5em}5\\
71 & skype.com & 2 & 0 & 0 & 1 & 0 & 1 & 0 & 1 & 0 & 0 & \hspace{.5em}5\\
72 & weather.com & 0 & 1 & 0 & 0 & 0 & 1 & 0 & 1 & 0 & 2 & \hspace{.5em}5\\

73 & bbc.com & 0 & 3 & 0 & 0 & 0 & 0 & 0 & 0 & 0 & 1 & \hspace{.5em}4\\
74 & cloudflare.com & 1 & 0 & 0 & 1 & 0 & 2 & 0 & 0 & 0 & 0 & \hspace{.5em}4\\
75 & fandom.com & 0 & 0 & 0 & 1 & 0 & 1 & 0 & 2 & 0 & 0 & \hspace{.5em}4\\
76 & mayoclinic.org & 0 & 1 & 0 & 0 & 0 & 1 & 0 & 0 & 0 & 2 & \hspace{.5em}4\\
77 & researchgate.net & 0 & 1 & 0 & 1 & 0 & 1 & 0 & 0 & 0 & 1 & \hspace{.5em}4\\
78 & slideshare.net & 0 & 1 & 0 & 0 & 0 & 1 & 0 & 0 & 2 & 0 & \hspace{.5em}4\\
79 & sourceforge.net & 2 & 0 & 0 & 1 & 0 & 0 & 0 & 0 & 1 & 0 & \hspace{.5em}4\\
\bottomrule
\end{tabular}}
\end{table*}}

\newcommand{\tableAllWebsitesPartTwo}[0]{
\begin{table*}[ht]
\centering
\scriptsize
\caption{The number of times each of the 158 websites appears in the top 50 of the 3 best and worst ranking nations, as determined by the GCI~\cite{index2020url} in each of the continents. Websites with the same total were ordered alphabetically. (continued)} 
\label{tab:allwebsitesPartTwo}
\rowcolors{3}{white}{lightgray!50}
\resizebox{.74\columnwidth}{!}{
\begin{tabular}{r p{9em}<{\raggedleft} | p{2.25em}<{\centering}p{2.25em}<{\centering} p{2.25em}<{\centering}p{2.25em}<{\centering} p{2.25em}<{\centering}p{2.25em}<{\centering} p{2.25em}<{\centering}p{2.25em}<{\centering} p{2.25em}<{\centering}p{2.25em}<{\centering} | c }
   \toprule
    && \multicolumn{2}{c}{\textbf{Africa}} & \multicolumn{2}{c}{\textbf{America}} & \multicolumn{2}{c}{\textbf{Asia}} & \multicolumn{2}{c}{\textbf{Europe}} & \multicolumn{2}{c|}{\textbf{Oceania}} \\
    \textbf{No.} &\multicolumn{1}{c|}{\textbf{Website}} & Best & Worst & Best & Worst & Best & Worst & Best & Worst & Best & Worst & \textbf{Total}\\
\midrule
80 & accuweather.com & 0 & 1 & 0 & 0 & 0 & 1 & 0 & 0 & 0 & 1 & 3\\
81 & bbc.co.uk & 1 & 0 & 0 & 1 & 0 & 0 & 0 & 0 & 1 & 0 & 3\\
82 & booking.com & 0 & 1 & 0 & 1 & 0 & 0 & 0 & 1 & 0 & 0 & 3\\
83 & gsmarena.com & 0 & 1 & 0 & 0 & 0 & 1 & 0 & 0 & 0 & 1 & 3\\
84 & godaddy.com & 2 & 0 & 0 & 0 & 0 & 0 & 0 & 0 & 1 & 0 & 3\\
85 & ilovepdf.com & 0 & 1 & 0 & 0 & 0 & 1 & 0 & 1 & 0 & 0 & 3\\
86 & macromedia.com & 0 & 0 & 1 & 0 & 1 & 0 & 1 & 0 & 0 & 0 & 3\\
87 & myshopify.com & 2 & 0 & 0 & 1 & 0 & 0 & 0 & 0 & 0 & 0 & 3\\
88 & samsung.com & 0 & 1 & 0 & 0 & 0 & 1 & 0 & 1 & 0 & 0 & 3\\
89 & savefrom.net & 0 & 2 & 0 & 0 & 0 & 1 & 0 & 0 & 0 & 0 & 3\\
90 & sciencedirect.com & 0 & 1 & 0 & 2 & 0 & 0 & 0 & 0 & 0 & 0 & 3\\
91 & telegram.org & 0 & 1 & 0 & 0 & 0 & 1 & 0 & 1 & 0 & 0 & 3\\
92 & washingtonpost.com & 1 & 0 & 0 & 0 & 0 & 0 & 0 & 0 & 2 & 0 & 3\\
93 & webmd.com & 0 & 0 & 0 & 0 & 0 & 1 & 0 & 0 & 0 & 2 & 3\\

94 & academia.edu & 0 & 1 & 0 & 0 & 0 & 1 & 0 & 0 & 0 & 0 & 2\\
95 & bongacams.com & 1 & 0 & 0 & 0 & 0 & 0 & 0 & 0 & 1 & 0 & 2\\
96 & dailymail.co.uk & 0 & 1 & 0 & 1 & 0 & 0 & 0 & 0 & 0 & 0 & 2\\
97 & discord.com & 0 & 0 & 0 & 1 & 0 & 0 & 0 & 1 & 0 & 0 & 2\\
98 & ecer.com & 0 & 1 & 0 & 0 & 0 & 0 & 0 & 0 & 0 & 1 & 2\\
99 & indeed.com & 0 & 1 & 0 & 1 & 0 & 0 & 0 & 0 & 0 & 0 & 2\\
100 & iqbroker.com & 0 & 1 & 0 & 0 & 0 & 0 & 0 & 0 & 0 & 1 & 2\\
101 & issuu.com & 0 & 0 & 0 & 0 & 0 & 0 & 0 & 1 & 1 & 0 & 2\\
102 & livemint.com & 0 & 0 & 0 & 0 & 0 & 0 & 0 & 0 & 0 & 1 & 2\\
103 & marca.com & 0 & 1 & 0 & 0 & 0 & 0 & 0 & 0 & 0 & 1 & 2\\
104 & medicalnewstoday.com & 0 & 1 & 0 & 0 & 0 & 0 & 0 & 0 & 0 & 1 & 2\\
105 & mega.nz & 0 & 1 & 0 & 0 & 0 & 0 & 0 & 1 & 0 & 0 & 2\\
106 & messenger.com & 0 & 0 & 0 & 0 & 0 & 0 & 0 & 0 & 0 & 2 & 2\\
107 & scribd.com & 0 & 1 & 0 & 0 & 0 & 1 & 0 & 0 & 0 & 0 & 2\\
108 & t.me & 0 & 1 & 0 & 0 & 0 & 0 & 0 & 0 & 0 & 1 & 2\\
109 & wetransfer.com & 0 & 1 & 0 & 0 & 0 & 0 & 0 & 1 & 0 & 0 & 2\\
110 & wikihow.com & 0 & 1 & 0 & 0 & 0 & 1 & 0 & 0 & 0 & 0 & 2\\
111 & wix.com & 0 & 0 & 0 & 0 & 0 & 0 & 0 & 0 & 2 & 0 & 2\\
112 & xinhuanet.com & 0 & 0 & 0 & 0 & 0 & 0 & 0 & 0 & 2 & 0 & 2\\
113 & ytmp3.cc & 0 & 0 & 0 & 0 & 0 & 0 & 0 & 0 & 0 & 2 & 2\\

114 & abc.net.au & 0 & 0 & 0 & 0 & 0 & 0 & 0 & 0 & 0 & 1 & 1\\
115 & as.com & 0 & 1 & 0 & 0 & 0 & 0 & 0 & 0 & 0 & 0 & 1\\
116 & bet365.com & 0 & 0 & 0 & 0 & 0 & 0 & 0 & 1 & 0 & 0 & 1\\
117 & britannica.com & 0 & 0 & 0 & 0 & 0 & 0 & 0 & 0 & 0 & 1 & 1\\
118 & businessinsider.com & 0 & 1 & 0 & 0 & 0 & 0 & 0 & 0 & 0 & 0 & 1\\
119 & cdc.gov & 0 & 0 & 0 & 0 & 0 & 0 & 0 & 0 & 0 & 1 & 1\\
120 & cnbc.com & 0 & 1 & 0 & 0 & 0 & 0 & 0 & 0 & 0 & 0 & 1\\
121 & cnblogs.com & 0 & 0 & 0 & 0 & 0 & 0 & 0 & 0 & 0 & 1 & 1\\
122 & deepl.com & 0 & 1 & 0 & 0 & 0 & 0 & 0 & 0 & 0 & 0 & 1\\
123 & detik.com & 0 & 0 & 0 & 0 & 0 & 1 & 0 & 0 & 0 & 0 & 1\\
124 & duckduckgo.com & 0 & 0 & 0 & 0 & 0 & 1 & 0 & 0 & 0 & 0 & 1\\
125 & elpais.com & 0 & 1 & 0 & 0 & 0 & 0 & 0 & 0 & 0 & 0 & 1\\
126 & eluniverso.com & 0 & 1 & 0 & 0 & 0 & 0 & 0 & 0 & 0 & 0 & 1\\
127 & genius.com & 0 & 0 & 0 & 0 & 0 & 0 & 0 & 0 & 0 & 1 & 1\\
128 & grid.id & 0 & 0 & 0 & 0 & 0 & 1 & 0 & 0 & 0 & 0 & 1\\
129 & hindustantimes.com & 0 & 0 & 0 & 0 & 0 & 0 & 0 & 0 & 0 & 1 & 1\\
130 & ikea.com & 0 & 0 & 0 & 0 & 0 & 0 & 0 & 1 & 0 & 0 & 1\\
131 & infobae.com & 0 & 1 & 0 & 0 & 0 & 0 & 0 & 0 & 0 & 0 & 1\\
132 & jianshu.com & 0 & 0 & 0 & 0 & 0 & 0 & 0 & 0 & 0 & 1 & 1\\
133 & kompas.com & 0 & 0 & 0 & 0 & 0 & 1 & 0 & 0 & 0 & 0 & 1\\
134 & kumparan.com & 0 & 0 & 0 & 0 & 0 & 1 & 0 & 0 & 0 & 0 & 1\\
135 & made-in-china.com & 0 & 0 & 0 & 0 & 0 & 0 & 0 & 0 & 0 & 1 & 1\\
136 & merdeka.com & 0 & 0 & 0 & 0 & 0 & 1 & 0 & 0 & 0 & 0 & 1\\
137 & naver.com & 1 & 0 & 0 & 0 & 0 & 0 & 0 & 0 & 0 & 0 & 1\\
138 & noodlemagazine.com & 0 & 0 & 0 & 0 & 0 & 1 & 0 & 0 & 0 & 0 & 1\\
139 & okezone.com & 0 & 0 & 0 & 0 & 0 & 1 & 0 & 0 & 0 & 0 & 1\\
140 & pikiran-rakyat.com & 0 & 0 & 0 & 0 & 0 & 1 & 0 & 0 & 0 & 0 & 1\\
141 & primevideo.com & 0 & 0 & 0 & 0 & 0 & 0 & 0 & 1 & 0 & 0 & 1\\
142 & proiezionidiborsa.it & 0 & 0 & 0 & 0 & 0 & 0 & 0 & 1 & 0 & 0 & 1\\
143 & remove.bg & 0 & 0 & 0 & 0 & 0 & 1 & 0 & 0 & 0 & 0 & 1\\
144 & shein.com & 0 & 1 & 0 & 0 & 0 & 0 & 0 & 0 & 0 & 0 & 1\\
145 & speedtest.net & 0 & 0 & 0 & 0 & 0 & 0 & 0 & 1 & 0 & 0 & 1\\
146 & suara.com & 0 & 0 & 0 & 0 & 0 & 1 & 0 & 0 & 0 & 0 & 1\\
147 & tencent.com & 0 & 0 & 0 & 0 & 0 & 0 & 0 & 0 & 0 & 1 & 1\\
148 & tokopedia.com & 0 & 0 & 0 & 0 & 0 & 1 & 0 & 0 & 0 & 0 & 1\\
149 & tribunnews.com & 0 & 0 & 0 & 0 & 0 & 1 & 0 & 0 & 0 & 0 & 1\\
150 & trustpilot.com & 0 & 0 & 0 & 0 & 0 & 0 & 0 & 1 & 0 & 0 & 1\\
151 & tokopedia.com & 0 & 0 & 0 & 0 & 0 & 1 & 0 & 0 & 0 & 0 & 1\\
152 & un.org & 0 & 1 & 0 & 0 & 0 & 0 & 0 & 0 & 0 & 0 & 1\\
152 & usgs.gov & 0 & 0 & 0 & 0 & 0 & 0 & 0 & 0 & 0 & 1 & 1\\
153 & usps.com & 0 & 0 & 0 & 0 & 0 & 0 & 0 & 0 & 0 & 1 & 1\\
154 & who.int & 0 & 0 & 0 & 0 & 0 & 0 & 0 & 0 & 1 & 0 & 1\\
155 & wikimedia.org & 0 & 1 & 0 & 0 & 0 & 0 & 0 & 0 & 0 & 0 & 1\\
156 & xe.com & 0 & 0 & 0 & 0 & 0 & 0 & 0 & 0 & 0 & 1 & 1\\
157 & y2mate.com & 0 & 1 & 0 & 0 & 0 & 0 & 0 & 0 & 0 & 0 & 1\\
158 & zara.com & 0 & 1 & 0 & 0 & 0 & 0 & 0 & 0 & 0 & 0 & 1\\
\bottomrule
\end{tabular}}
\end{table*}}
\definecolor{bestColor}{HTML}{009E73}
\definecolor{worstColor}{HTML}{E69F00}

\newcommand{\figCompromiseDiscovery}[0]{
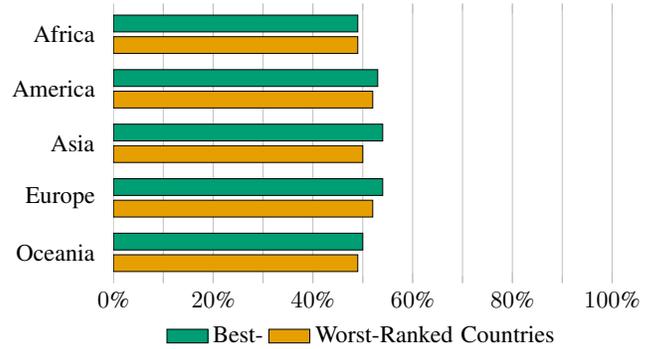
\begin{figure}[t]
\centering
\resizebox{.975\columnwidth}{!}{
    \begin{tikzpicture}
        \begin{axis}[
        width=\columnwidth,
        xbar,
        bar width = 7pt,
        xmin=0,
        xmax=100,
        area legend,
        reverse legend,
        legend style={
        legend columns=2,
            at={(xticklabel cs:0.5)},
            anchor=north,
            draw=none
        },
        xticklabel={$\pgfmathprintnumber{\tick}$\%},
     	y=8mm, 
        enlarge y limits={abs=0.45cm}, 
     	every tick/.style={very thin, color=lightgray}, 
        xmajorgrids = true, 
        xminorgrids = true, 
     	minor tick num=1, 
        y axis line style = { opacity = 0 }, 
        ytick style = {draw=none}, 
        symbolic y coords = {
            {Oceania},
            {Europe},
            {Asia},
            {America},
            {Africa}},
        ytick=data, 
        ]
        \addplot[fill=worstColor] coordinates {
            (49,{Oceania}) 
            (52,{Europe}) 
            (50,{Asia}) 
            (52,{America}) 
            (49,{Africa})}; 
        \addplot[fill=bestColor] coordinates {
            (50,{Oceania}) 
            (54,{Europe}) 
            (54,{Asia}) 
            (53,{America}) 
            (49,{Africa})}; 
        \legend{Worst-Ranked Countries, Best-}
        \end{axis}
    \end{tikzpicture}}
\caption{Average presence of advice for \textit{Phase 1: Compromise Discovery} in each of the 5 analyzed continents.}
\label{fig:compromiseDiscovery}
\end{figure}}

\newcommand{\figAccountRecovery}[0]{
\begin{figure}[t]
\centering
\resizebox{.975\columnwidth}{!}{
    \begin{tikzpicture}
        \begin{axis}[
        width=\columnwidth,
        xbar,
        bar width = 7pt,
        xmin=0,
        xmax=100,
        area legend,
        reverse legend,
        legend style={
        legend columns=2,
            at={(xticklabel cs:0.5)},
            anchor=north,
            draw=none
        },
        xticklabel={$\pgfmathprintnumber{\tick}$\%},
     	y=8mm, 
        enlarge y limits={abs=0.45cm}, 
     	every tick/.style={very thin, color=lightgray}, 
        xmajorgrids = true, 
        xminorgrids = true, 
     	minor tick num=1, 
        y axis line style = { opacity = 0 }, 
        ytick style = {draw=none}, 
        symbolic y coords = {
            {Oceania},
            {Europe},
            {Asia},
            {America},
            {Africa}},
        ytick=data, 
        ]
        \addplot[fill=worstColor] coordinates {
            (71,{Oceania}) 
            (73,{Europe}) 
            (70,{Asia}) 
            (72,{America}) 
            (69,{Africa})}; 
        \addplot[fill=bestColor] coordinates {
            (71,{Oceania}) 
            (77,{Europe}) 
            (77,{Asia}) 
            (75,{America}) 
            (72,{Africa})}; 
        \legend{Worst-Ranked Countries, Best-}
        \end{axis}
    \end{tikzpicture}}
\caption{Average presence of advice for  \textit{Phase 2: Account Recovery} in each of the 5 analyzed continents.}
\label{fig:accountRecovery}
\end{figure}
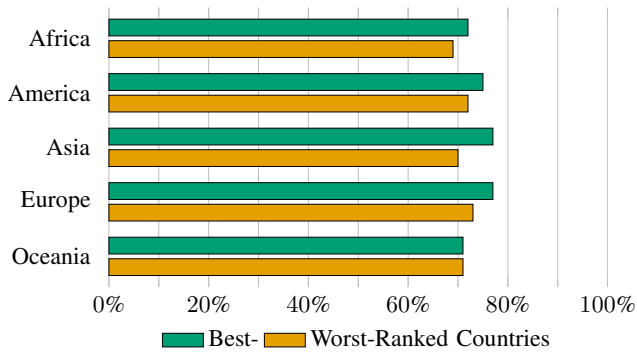}

\newcommand{\figLimitAccess}[0]{
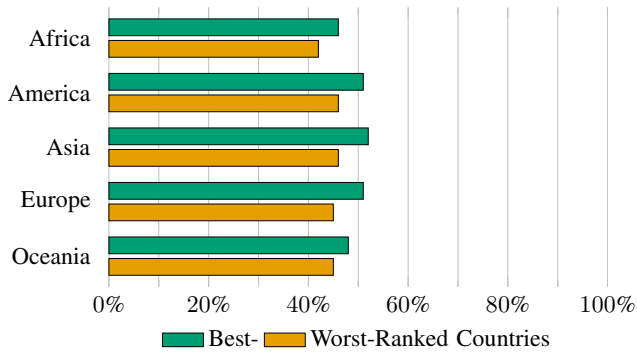
\begin{figure}[t]
\centering
\resizebox{.975\columnwidth}{!}{
    \begin{tikzpicture}
        \begin{axis}[
        width=\columnwidth,
        xbar,
        bar width = 7pt,
        xmin=0,
        xmax=100,
        area legend,
        reverse legend,
        legend style={
        legend columns=2,
            at={(xticklabel cs:0.5)},
            anchor=north,
            draw=none
        },
        xticklabel={$\pgfmathprintnumber{\tick}$\%},
     	y=8mm, 
        enlarge y limits={abs=0.45cm}, 
     	every tick/.style={very thin, color=lightgray}, 
        xmajorgrids = true, 
        xminorgrids = true, 
     	minor tick num=1, 
        y axis line style = { opacity = 0 }, 
        ytick style = {draw=none}, 
        symbolic y coords = {
            {Oceania},
            {Europe},
            {Asia},
            {America},
            {Africa}},
        ytick=data, 
        ]
        \addplot[fill=worstColor] coordinates {
            (45,{Oceania}) 
            (45,{Europe}) 
            (46,{Asia}) 
            (46,{America}) 
            (42,{Africa})}; 
        \addplot[fill=bestColor] coordinates {
            (48,{Oceania}) 
            (51,{Europe}) 
            (52,{Asia}) 
            (51,{America}) 
            (46,{Africa})}; 
        \legend{Worst-Ranked Countries, Best-}
        \end{axis}
    \end{tikzpicture}}
\caption{Average presence of advice for \textit{Phase 3: Limit Unwanted Access} in each of the 5 analyzed continents.}
\label{fig:limitAccess}
\end{figure}}

\newcommand{\figServiceRestoration}[0]{
\begin{figure}[t]
\centering
\resizebox{.975\columnwidth}{!}{
    \begin{tikzpicture}
        \begin{axis}[
        width=\columnwidth,
        xbar,
        bar width = 7pt,
        xmin=0,
        xmax=100,
        area legend,
        reverse legend,
        legend style={
        legend columns=2,
            at={(xticklabel cs:0.5)},
            anchor=north,
            draw=none
        },
        xticklabel={$\pgfmathprintnumber{\tick}$\%},
     	y=8mm, 
        enlarge y limits={abs=0.45cm}, 
     	every tick/.style={very thin, color=lightgray}, 
        xmajorgrids = true, 
        xminorgrids = true, 
     	minor tick num=1, 
        y axis line style = { opacity = 0 }, 
        ytick style = {draw=none}, 
        symbolic y coords = {
            {Oceania},
            {Europe},
            {Asia},
            {America},
            {Africa}},
        ytick=data, 
        ]
        \addplot[fill=worstColor] coordinates {
            (52,{Oceania}) 
            (41,{Europe}) 
            (40,{Asia}) 
            (42,{America}) 
            (44,{Africa})}; 
        \addplot[fill=bestColor] coordinates {
            (43,{Oceania}) 
            (45,{Europe}) 
            (46,{Asia}) 
            (45,{America}) 
            (43,{Africa})}; 
        \legend{Worst-Ranked Countries, Best-}
        \end{axis}
    \end{tikzpicture}}
\caption{Average presence of advice for \textit{Phase 4: Service Restoration} in each of the 5 analyzed continents.}
\label{fig:serviceRestoration}
\end{figure}
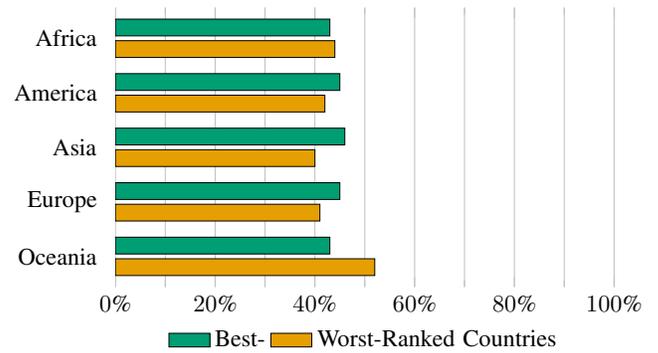}

\newcommand{\figPrevention}[0]{
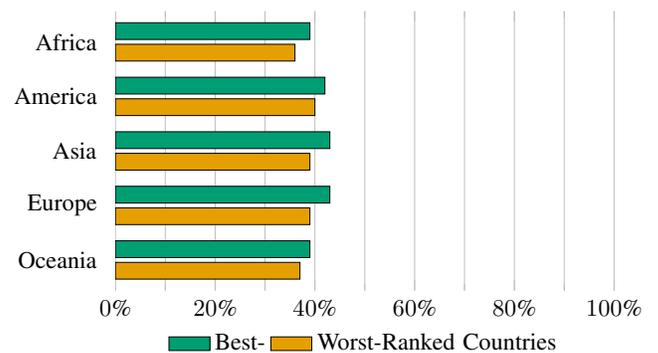
\begin{figure}[t]
\centering
\resizebox{.975\columnwidth}{!}{
    \begin{tikzpicture}
        \begin{axis}[
        width=\columnwidth,
        xbar,
        bar width = 7pt,
        xmin=0,
        xmax=100,
        area legend,
        reverse legend,
        legend style={
        legend columns=2,
            at={(xticklabel cs:0.5)},
            anchor=north,
            draw=none
        },
        xticklabel={$\pgfmathprintnumber{\tick}$\%},
     	y=8mm, 
        enlarge y limits={abs=0.45cm}, 
     	every tick/.style={very thin, color=lightgray}, 
        xmajorgrids = true, 
        xminorgrids = true, 
     	minor tick num=1, 
        y axis line style = { opacity = 0 }, 
        ytick style = {draw=none}, 
        symbolic y coords = {
            {Oceania},
            {Europe},
            {Asia},
            {America},
            {Africa}},
        ytick=data, 
        ]
        \addplot[fill=worstColor] coordinates {
            (37,{Oceania}) 
            (39,{Europe}) 
            (39,{Asia}) 
            (40,{America}) 
            (36,{Africa})}; 
        \addplot[fill=bestColor] coordinates {
            (39,{Oceania}) 
            (43,{Europe}) 
            (43,{Asia}) 
            (42,{America}) 
            (39,{Africa})}; 
        \legend{Worst-Ranked Countries, Best-}
        \end{axis}
    \end{tikzpicture}}
\caption{Average presence of advice for \textit{Phase 5: Prevention} in each of the 5 analyzed continents.}
\label{fig:prevention}
\end{figure}}

\newcommand{\figOverall}[0]{
\begin{figure}[t]
\centering
\resizebox{.975\columnwidth}{!}{
    \begin{tikzpicture}
        \begin{axis}[
        width=\columnwidth,
        xbar,
        bar width = 7pt,
        xmin=0,
        xmax=100,
        area legend,
        reverse legend,
        legend style={
        legend columns=2,
            at={(xticklabel cs:0.5)},
            anchor=north,
            draw=none
        },
        xticklabel={$\pgfmathprintnumber{\tick}$\%},
     	y=8mm, 
        enlarge y limits={abs=0.45cm}, 
     	every tick/.style={very thin, color=lightgray}, 
        xmajorgrids = true, 
        xminorgrids = true, 
     	minor tick num=1, 
        y axis line style = { opacity = 0 }, 
        ytick style = {draw=none}, 
        symbolic y coords = {
            {Oceania},
            {Europe},
            {Asia},
            {America},
            {Africa}},
        ytick=data, 
        ]
        \addplot[fill=worstColor] coordinates {
            (48,{Oceania}) 
            (47,{Europe}) 
            (46,{Asia}) 
            (47,{America}) 
            (45,{Africa})}; 
        \addplot[fill=bestColor] coordinates {
            (47,{Oceania}) 
            (51,{Europe}) 
            (51,{Asia}) 
            (50,{America}) 
            (47,{Africa})}; 
        \legend{Worst-Ranked Countries, Best-}
        \end{axis}
    \end{tikzpicture}}
\caption{Average presence of advice across all 5 phases in each of the 5 analyzed continents.}
\label{fig:overview}
\end{figure}
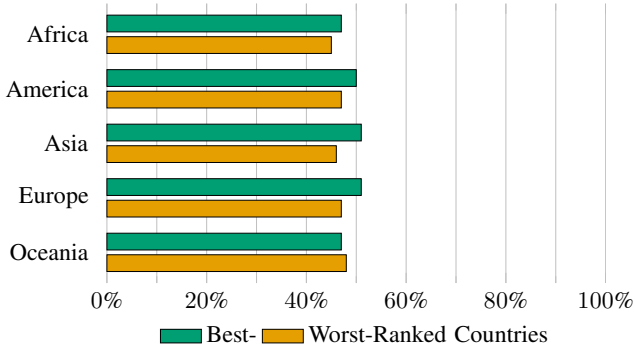}

\begin{abstract}
Websites are used regularly in our day-to-day lives, yet research has shown that it is challenging for many users to use them securely, e.g., most prominently due to weak passwords through which they access their accounts. At the same time, many services employ low-security measures, making their users even more prone to account compromises with little to no means of remediating compromised accounts. Additionally, remediating compromised accounts requires users to complete a series of steps, ideally all provided and explained by the service. However, for U.S.-based websites, prior research has shown that the advice provided by many services is often incomplete. To further understand the underlying issue and its implications, this paper reports on a study that analyzes the account remediation procedure covering the $50$ most popular websites in $30$ countries,~$6$ each in Africa, the Americas, Asia, Europe, and Oceania. We conducted the first transcontinental analysis on the account remediation protocols of popular websites. The analysis is based on $5$ steps websites need to provide advice for: compromise discovery, account recovery, access limitation, service restoration, and prevention.
We find that the lack of advice prior work identified for websites from the U.S. also holds across continents, with the presence ranging from 37\% to 77\% on average. Additionally, we identified considerable differences when comparing countries and continents, with countries in Africa and Oceania significantly more affected by the lack of advice. To address this, we suggest providing publicly available and easy-to-follow remediation advice for users and guidance for website providers so they can provide all the necessary information. 
\end{abstract}

\section{Introduction}
We use websites in everyday life for various purposes, including necessities of life, healthcare, education, business, and information dissemination~\cite{wohlmacher2000applications}. Several of these websites require users to create an account to access the content, which causes them to include several individual details~\cite{jammalamadaka2006delegate}. Hence, accounts can hold sensitive and critical information of users, some of which is related to health, social affairs, or political affairs~\cite{josang2007security}. Moreover, technological capabilities and the wealth of data stored through these websites make it lucrative for attackers to steal, destroy, or modify data. Research has shown that every~39 second, a website is attacked, and on average, 30,000 websites are attacked every day. Furthermore, such attacks on websites increase at a rate of 13\% each year and can be executed in various ways~\cite{costigan2021sovereign}. 

Due to the described risks, websites can become insecure and easy to compromise, out of which the user accounts become a primary target. 
Once an account is compromised, it must be restored to its pre-compromise state, referred to as ``account remediation''~\cite{neil-21-acc-remediation-adv}. 
In general, account remediation is a set of protocols to transform a compromised account into one that is again entirely under the user's control. 
This process includes several aspects, e.g., the mechanisms of resetting the password, turning on two-factor authentication, security verification checks, or account deletion~\cite{neil-21-acc-remediation-adv}. 
Hence, it can be technically complex and, in several cases, is left to user choices and discretion~\cite{mcclelland2001failures,wu2015risk}. Thus, lacking information about account remediation is one of the most critical factors that often keep users un/misinformed about actions to perform if an account is compromised. Moreover, users depend on a specific provision of advice by the website owners to protect their accounts from hacks. Unfortunately, prior research has shown that among U.S.-based websites, these instructions are often incomplete, making it cumbersome or even impossible for users to complete the process~\cite {neil-21-acc-remediation-adv}. 

\vspace{-.1em}

To further explore this, we report on an analysis covering the remediation advice of the $50$ most popular websites in $30$ countries across $5$ continents, Africa, America, Asia, Europe, and Oceania. First, we selected $6$ countries from each continent based on each country's ranking in Global Cybersecurity Index (GCI) 2020~\cite{index2020url}, the $3$ countries with the best and $3$ with the worst rating. Afterward, we assembled a list of the $50$ most popular websites in each country based on the Tranco list~\cite{lepochat-19-tranco}. Several of these websites are popular across countries; hence, the final list for the analysis consisted of $158$ unique websites. In the analysis, we checked each website for the presence of account remediation advice using an updated model from prior work~\cite{neil-21-acc-remediation-adv}, which defines account remediation in 5 phases: \textit{Compromise Discovery}, \textit{Account Recovery}, \textit{Limit Access}, \textit{Service Restoration}, and \textit{Prevention}. Finally,  we compared the $5$ different continents based on the analysis results.

\vspace{-.1em}

We believe automation of all possible steps is vital to unburdening users and limiting an account compromise's adverse effects. Consequently, the research object was to analyze the current remediation process to identify missing, redundant, or unclear steps on a global scale. Ultimately, through this research and with the future extension of this work, we aim to create an account remediation protocol that is optimized both regarding maximizing the technical assistance and the guidance of steps where such assistance is not possible. Through our analysis, we noted the following issues in the account remediation protocols of websites popular in different places of the world, which is the contribution of this work:

\begin{itemize}

 \item \textit{Lack of Information}: We found that most websites have pages with policies, including the privacy policy. However, they often lack instructions, e.g., on creating a strong password or deleting an account, which is similar to results from prior work~\cite{adhikari2022privacy,das2019privacy}. In addition, websites should address their protection for accounts, the procedures the user must follow if the account was hacked, or the user's desire to restore the account and its content.

 \item \textit{Lack of Security Measures}: Most websites do not use multiple security checks to verify the user's authentication. For example, they do not offer two-factor authentication which is recommended by security experts~\cite{das2020risk,jensen2021multi,zhang2022building} and are limited to verification only by sending an email to the user. Frequently, they also do not check if a newly set password is related to the one being replaced---a test possible without any risk as knowledge of the old password also needs to be checked. 
 
 \item \textit{Difference Between Countries}: We found that remediation advice needs to be added on websites popular across continents. However, users in countries with lower security standards are more prone to needing more advice. While most of these differences appear manageable, users in countries that are part of the global south in Africa and Oceania are significantly more affected by the absence of crucial remediation advice. We want to mention that in this work, we are intentionally not using phrases such as ``developed and developing nations'' or ``Third World''; instead, we are using the terms ``Global South'' and ``Global North.'' These phrases shift from a focus on development or economic differences to an emphasis on geopolitical power relations~\cite{dados2012global}. Moreover, prior work suggests that the concepts or phrasing such as ``Third World'' is no longer a viable concept from a political perspective and can be patronizing~\cite{kolko2009computer,murphy2008economic}. Nevertheless, we understand that these terms may still inspire similar discussions, which can be presently represented with the phrasing of the global south and global north into a force in the reconfiguration of global relations~\cite{dirlik2007global}.

 \item \textit{Identification of the Impact of Third Parties on Account Remediation}: 
 The presence of third parties means data is shared between services for commercial purposes, threatening the users' privacy and the security of their accounts. 
 Some websites name third parties they share data with, but many websites say they send data to third parties but do not provide easy-to-find and sometimes any information, which parties particularly.
 
 \item \textit{Ignorance of Recommendations From Research}: Numerous recommendations have evolved from research in the field of (usable) security throughout the years; still, it depends on the website to also adopt them~\cite{fagan2018follow}. We noted that these highly used websites must implement several advice pieces. For example, 2FA is a recommendation by experts to protect user accounts~\cite{reeder-17-152-simple-steps,shirvanian20212d}, yet, most websites do not offer 2FA at all, and those which do often only send one-time passwords (OTP) via email. 
  
\end{itemize}

In Section~\ref{sec:related}, we detail related work focusing on how the account compromise and remediation protocols came into existence and how users become the primary actors in making these protocols a success. Section~\ref{sec:method} depicts the process of collecting the websites and developing the codebook to analyze the account remediation procedures of websites. Next, Section~\ref {sec:results} depicts an overview of the collected data and the analysis across different continents and countries. Section~\ref{sec:discussion} discusses and contextualizes the results, which forms the basis for our recommendation in Section~\ref{sec:implications}. Finally, we provide an overview of the limitations of this study, which we plan to address through the future extension of this work in Section~\ref{sec:limit}, and conclude the paper in Section~\ref{sec:conclusion}.
\section{Related Work}
\label{sec:related}

Account remediation is a critical process for online interaction to protect user data; however, it can be complicated. 
For our analysis, we selected the account remediation protocol developed by Neil et al.~\cite{neil-21-acc-remediation-adv} to understand the account remediation aspects. They investigated the account remediation procedures of $57$ U.S.-based websites and identified 5 phases that compose the process: 
\begin{enumerate*}[label=(\arabic*)]
    \item compromise discovery,
    \item account recovery, 
    \item limit access,
    \item service restoration, and
    \item prevention.
\end{enumerate*}
We extended their methodology for a transcontinental analysis of $158$ websites covering the top $50$ websites visited in 5 continents and across 6 countries from each continent in combination to get a broader understanding of the protocol impacts and application for account remediation.
Below we also want to address the work in the 3 most related research areas: discovering an account compromise, risk mitigation, and users' decision-making process. 

\subsection{Account Compromise Discovery}
Researchers have emphasized account compromise discovery to improve the timeliness and ability of websites to effectively detect compromised accounts~\cite{egele-13-compa,egele-15-towards-detecting,halawa-17-an-early,igawa-15-recognition-of,li-19-checking-credentials}. 
Additional work on prevention mechanisms drew attention to best practices that can mitigate the risk of account compromise, e.g., a risk analysis~\cite{bursztein-14-handcrafted-fraud,walsh-21-intercultural-analysis}. 
With our study, we focus on the general protocols provided by the websites. 

Halawa et al. suggested an early warning system based on machine learning to detect compromised and vulnerable accounts by identifying suspicious account usage behavioral patterns~\cite{halawa-17-an-early}. Egele et al. have also leveraged machine learning techniques to detect compromised accounts in social networks by introducing a tool called \textit{COMPA}. COMPA creates behavioral profiles for Facebook and Twitter users, trains a model with a small manually labeled dataset of compromised and non-compromised user accounts, and uses this model to detect compromised Twitter and Facebook accounts~\cite{egele-13-compa}.

\subsection{Risk Mitigation}
Work on prevention mechanisms draws attention to best practices that can mitigate the risk of account compromise— Prior researchers advocate for defense strategies such as using second-factor authentication~\cite{bursztein-14-handcrafted-fraud,das2019security,das2020mfa,das2019evaluating}. Doerfler et al., on the other hand, investigated the effectiveness of these prevention mechanisms and their impact on users~\cite{doerfler-19-login-challenges}. 
Other research analyzed different fallback authentication mechanisms that can serve as tools for users that have lost access to their accounts~\cite{bonneau-12-the-quest, markert-19-comparative-study}. These forms which have been analyzed, both in terms of security and usability, can be separated into 4 different types: email-based resets~\cite{garfinkel-03-email}, SMS-based systems~\cite{bonneau-15-lies-and-account-recovery}, personal knowledge questions~\cite{just-09-pkq-usability,pinchot-12-facebook-pkq,rabkin-08-pkq-facebook,schechter-09-no-secret}, and social authentication measures~\cite{brainard-06-social, schechter-09-social}. 

Of all those methods, Li et al. identified the email-based reset as the most popular~\cite{li-18-email-master-key}. Furthermore, they propose an improved method called \textit{Secure Email Account Recovery} to prevent malicious activities when the registered email account is compromised. However, in addition to the technical implementation, it is also essential to understand the user perspective. At the same time, expert advice is valued for privacy, cybersecurity auditing, and research. Nevertheless, since advice also needs to be provided at the appropriate places, the utilization of websites and applications is as important~\cite{redmiles-16-advice-digital-security}. Through our study, we intend to explore and check this.

\subsection{Users' Decision-Making Process}
Along these lines, regarding the decision-making procedures of account remediation and online account compromise reactions, researchers have also focused on users' mental models~\cite{bravo-lillo-11-warning-mental,kelsey-19-entertainment-media}.
While discussing decision-making procedures of account remediation and online account compromise reactions, researchers have also focused on mental models in the face of a hacked account. The key finding was that understanding security measures is incomplete~\cite{rader-12-informal-lessons,shay-14-religious-aunt}. 

Zangerle and Specht further analyzed the user behavior after a discovered account compromise~\cite{zangerle-14-sorry-hacked}. 
They found that 27.3\% of users whose accounts were exposed chose to change to a new account. This result is interesting as it sees users possibly adding or circumventing a new category in account remediation---choosing to create a new account entirely instead of relying on the account remediation process to regain control and prevent future compromises.
\label{sec:relatedwork}
\section{Method}
\label{sec:method}

This section describes the method used for the transcontinental analysis of account remediation protocols. We collected the top 50 websites from 6 countries and across 5 continents worldwide to conduct our evaluation of the account remediation protocol. We intended to get a holistic overview of the account remediation protocols adapted by several websites. Therefore, our study design includes the following 2 steps: 
First, we describe the collection of the relevant websites in Section~\ref{sec:websites}, which we composed such that they represent the most popular websites in each of the continents. Followed by that, Section~\ref{sec:code} depicts how we updated the codebook for our analysis, which was initially developed by Neil et al.~\cite{neil-21-acc-remediation-adv}.

\tableCountries{}

\subsection{Website Collection}
\label{sec:websites}
Previous work~\cite{neil-21-acc-remediation-adv} found that the information about account remediation procedures on U.S.-based websites is oftentimes incomplete. Our study aims to evaluate the account remediation for 50 websites from 6 countries on 5 continents globally. 
This was done to provide an overview of how the different websites in different geographical locations implement expert-suggested account remediation protocols. 
The purpose of the analysis is to gain an in-depth evaluation of those websites in several areas for account remediation. 

To achieve this, we took 5 continents, Africa, America (North and South), Asia, Europe, and Oceania.
We eliminated Antarctica from the analysis, given its low population.
To determine the 3 best and worst countries on each continent regarding their cybersecurity proficiency, we used the Global Cybersecurity Index (GCI)~\cite{index2020url}.
Note that the GCI combines South and North America; hence, we also kept them together for the analysis~\cite{bruggemann2022global}. 
We chose GCI as a reference since it assesses nations through the expertise of different organizations across legislative measures, technical measures, organizational measures, capacity development, and collaboration - and then aggregates this assessment into an overall score. 

The cybersecurity proficiency of a country is estimated based on the ability to protect its infrastructure, including systems, applications, hardware, software, and data. 
The GCI also evaluates a country's legal measures, such as legislation and laws; technical measures, such as the deployment of incident response teams; organizational measures that examine the governance and coordination mechanisms addressing cybersecurity within a country; capacity development measures to counteract large-scale attacks, cooperative measures among multiple stakeholders or countries, and child protection measures.
Taking the GCI ranking, we ended up with 30 countries, the 3 best and 3 worst for each continent, as depicted in Table~\ref{tab:countries}.

After selecting the countries, we used the Tranco list~\cite{lepochat-19-tranco} to identify the 50 most popular websites for each country. The top websites are calculated based on the number of people who visit the website per month, as listed in Tranco's List. The list was verified and calculated based on a combination of Alexa, Umbrella, and Majestic websites. The second ranking to classify the websites was the ``Alexa rank,'' a global popularity ranking~\cite{thakur2011quantitative}. It uses web traffic data to make an ordered list of each country's most popular sites on the Internet. We used multiple categories and rankings to ensure we got the most visited and searched websites for the selected countries. 
By doing this, we ended up with a list of 1394 websites, $300$ each from America, Europe, and Asia, $253$ websites from Africa, and $241$ websites from Oceania. The numbers being lower for Africa and Oceania is due to the sparsity of data. Hence, the Tranco list for Eritrea, the Marshall Islands, the Solomon Islands, and Vanuatu consists of less than $50$ websites. 
We merged the lists from all nations and deleted duplicates, yielding a list of $158$ unique websites. They are depicted in Table~\ref{tab:allwebsitesPartOne} and \ref{tab:allwebsitesPartTwo} along with their number of occurrences in the top 50 list in each country.

\subsection{Codebook Composition}
\label{sec:code}
A first step toward analyzing the account remediation procedure on websites was done by Neil et al. 
They developed a codebook with a total of 32 codes and identified 5 phases they can be grouped into: 
\begin{enumerate*}[label=(\arabic*)]
    \item compromise discovery,
    \item account recovery, 
    \item limit access,
    \item service restoration, 
    \item prevention.
\end{enumerate*}
The first phase is where the account owner identifies the compromise to initiate the remediation procedure. 
Afterward, access to the account may need to be restored, e.g., if the attacker changed the password.
The next step is to remove any malicious access to the account, e.g., by stopping unknown sessions. 
Step 4 is to identify and restore any unwanted changes, followed by step 5, which aims at reducing the chances of future compromise by following a set of security advice. 
By completing all 5 phases, a compromised account becomes again an account entirely under control by the actual account owner.

To analyze the account remediation protocols of the~$158$ websites we collected based on the method described in Section~\ref{sec:websites}, we first updated the codebook by removing and adding 3 codes.
We then distributed the websites among 5 coders, who used the updated codebook to examine them. 
To ensure the quality and correctness of the data, we performed a cross-check for a randomly selected $10\%$ of the websites. 

The first 2 codes we removed are \textit{Run Endpoint Security} and \textit{Run Endpoint Security Solutions}. They are part of the second (\textit{Account Recovery}) and fifth (\textit{Prevention}) phases, respectively, and advise users to run their antivirus software to prevent any potential attacks. 
We removed both codes as the benefit of such software on computers is disputable~\cite{ion-15-hack-my-mind,mihalcik-20-antivirus-privacy, temperton-15-antivirus-privacy} and on mobile devices with their dedicated sandbox environments and app stores, the situation is similar~\cite{av-19-antivirus-test}. 
The third code we removed is \textit{Physical Security} which is part of phase 5 \textit{Prevention} and intends to prevent undesired access by strangers. 
However, it is a code that was barely found in prior work~\cite{neil-21-acc-remediation-adv}, and it only applies to specific use cases, e.g., when using a computer in a café. On the other hand, mobile devices, presumably the most used public, are usually locked after usage~\cite{harbach-16-the-anatomy, harbach-16-keep-on,harbach-14-hard-lock-life}. 

The first code we added is \textit{Password Rotation} which became part of phase 2 \textit{Account Recovery}. It describes if the website informs the user (and assures) that when changing the password, it must not be related to the old one. If this check is missing, an attacker with credentials may still be able to use them even though the password itself was changed~\cite{bhagavatula-20-change-after-breach}. 
The following code, added to the third phase \textit{Limit access}, is \textit{Instructions for Account Deletion}. Deleting the account itself is also a way to limit access, primarily if users have not used the account for a while and realize they no longer need it~\cite{zangerle-14-sorry-hacked}. 
Finally, we added a code \textit{Regular Security Checkups/Advice} to phase 5 \textit{Prevention}. It depicts if websites inform their users to check the security of their accounts regularly because once the account is remediated, users may shift their focus away from security aspects which is why this long-term advice is essential. Furthermore, to also cover the modality in which certain functionalities are offered, we additionally collected information about the communication channel for notifications (email, SMS, push) and how users can contact customer support (email, chat, phone, form).

\tableAnalysisCombined{}
\section{Results}
\label{sec:results}

We now present the results for the analysis of the $158$ unique websites, representing the top $50$ websites for each of the $6$ countries in Africa, America, Asia, Europe, and Oceania. 
The full details for each continent's best and worst countries are presented in Table~\ref{tab:analysis}, where each entry reflects the percentage of websites with established advice on a given account remediation phase. In addition, the averages for each phase are shown in Figures~\ref{fig:compromiseDiscovery}--\ref{fig:prevention}, and an accumulated overview is given in Figure~\ref{fig:overview}. To compare the presence of account remediation advice across different groups, we used a $\chi^2$ test ($\alpha = 0.05$) with the Benjamini-Hochberg procedure applied for posthoc pairwise comparisons~\cite{lassak2021s,lee2018proper,zimmermann2022hybrid}.

\figCompromiseDiscovery{}

\subsection*{Phase 1: Compromise Discovery}
\label{sec:compromiseDiscovery}

The first phase, in which a user must detect the compromise, consists of 9 steps. As can be seen in Figure~\ref{fig:compromiseDiscovery}, about half of the required advice is present across all continents and countries. Most advice (54\%) is on the most popular websites for the highest-ranked countries in Europe and Asia. Conversely, the least advice (49\%) is given on websites that are popular in Africa (best and worst) or the Marshall Islands, Solomon Islands, and Vanuatu, i.e., the lowest-ranked countries according to the GCI~\cite{index2020url} in Oceania. 

The one type of advice on most websites, irrespective of the continent, is \textit{Explicit Notification}, i.e., websites notify users about a new login. Interestingly, the highest number appears in America (91\%) for the popular websites of the lowest-ranked countries according to the GCI~\cite{index2020url} in the form of Honduras, Dominica, and Haiti. Conversely, this number goes down to 73\% for websites popular in the worst-ranked countries in Oceania (Marshall Islands, Solomon Islands, and Vanuatu). The $\chi^2$ test also suggested some significant differences in the data for this code ($p=0.023$). However, the post hoc analysis did not confirm this initial assumption. 

We observe the lowest scores for this first phase for websites informing users about a connected social media or third-party account. The percentages range from 18\% for websites popular in the worst-ranked countries in Europe to 38\% for popular websites in Oceania, again the lowest-rated countries. While this finding may seem counterintuitive, it could be partially explained by the overall lower popularity of this feature across websites popular in the Marshall Islands, Solomon Islands, and Vanuatu. Hence, the few websites that give this form of advice, such as Facebook, Instagram, and LinkedIn, stand out more. Finally, we could not confirm the significance of these differences ($p=0.115$).

We observed significant differences for \textit{Billing/Finance Issues} where websites inform their users, e.g., about irregular charges. Percentages range from 64\% for websites popular in the best-ranked countries in Europe to 32\% for websites in the lowest-ranked countries in Oceania. The difference between these 2 groups is significant ($p < 0.05$), as is the difference between the best-ranked countries in America (63\%) and Asia (63\%) compared to the lowest-ranked countries in Africa (43\%) and Oceania~(32\%). While this difference is problematic in itself, the fact that it is reflected between high-income and low-income countries further adds to the seriousness of the problem. We will further discuss this issue, which we observed throughout all 5 phases in Section~\ref{sec:discussion}.

\subsection*{Phase 2: Account Recovery}
\label{sec:accountRecovery}
The average presence of advice for the \textit{Account Recovery} phase is shown in Figure~\ref{fig:accountRecovery}. Overall, the percentages are the highest across all 5 steps, ranging from 69\% for the worst-ranked countries in Africa to 77\% for the best-ranked countries in Asia and Europe. In addition, popular websites in the best-ranked countries score higher, or at least identical in the case of Oceania, compared to the worst-ranked countries. 

A more in-depth analysis based on the results depicted in Table~\ref{tab:analysis} shows that \textit{Customer Service Process} and \textit{Password Reset} account for the overall high presence of advice. Both range from $\sim$80\% to $\sim$90\%, showing that most websites instruct their users to change their passwords and contact them in case of difficulties. In contrast, many services do not check if users try to use their potentially breached password again, nullifying any security increase. Numbers range from 38\% for websites popular in the worst-ranked countries in Africa and Oceania to 58\% for the best-ranked countries in Asia and Europe. Generally, this check is less prevalent across the worst-ranked countries than their best-ranked counterparts on each continent. The initial $\chi^2$ test confirmed this observation ($p < 0.05$), yet we could not observe any actual significance for the pairwise comparisons. 

\figAccountRecovery{}
\figLimitAccess{}
\figServiceRestoration{}

\subsection*{Phase 3: Limit Unwanted Access}
\label{sec:limitAccess}
The third phase of account remediation intends to limit unwanted access. Here the average presence of advice depicted in Figure~\ref{fig:limitAccess} ranks from 42\% for the most popular website in Equatorial Guinea, Eritrea, and Burundi, i.e., the worst-ranked countries in Africa, to 52\% for websites popular in the best-ranked countries in Asia. Nevertheless, advice is more present across popular websites in better-ranked countries, up to a difference of 6\% between the 2 groups of countries in Asia (52\% vs. 46\%) and Europe (51\% vs. 45\%). 

Regarding the 5 codes for this phase, \textit{Instructions for Account Deletion} is the most popular, ranging from 70\% to~87\%. For this code, the $\chi^2$ test also revealed some significance ($p < 0.01$); however, the subsequent pairwise tests did not confirm this. Advice from this phase which is less present on websites, is the instruction to sign out everywhere to ensure no malicious session stays active. In the worst-ranked countries in Africa, only 19\% of the website provide this advice; for better-ranked countries, this goes up to 29\%. We observed a similar presence for advice telling users to log out from unknown sessions, which is a less invasive way of dealing with the problem. However, it makes users responsible for the complex task of telling malicious and benign sessions apart. 

\figPrevention{}

\subsection*{Phase 4: Service Restoration}
\label{sec:serviceRestoration}
Phase 4~\textit{Service Restoration} provides users with instructions on restoring their account's initial status. As seen in Figure~\ref{fig:serviceRestoration}, advice in this phase is given by 40\% to 46\% of the websites with one exception: the worst-ranked countries in Oceania, the Marshall Islands, Solomon Islands, and Vanuatu. This stark contrast of 52\%, especially in comparison to the previous analysis, can be traced back to a comparatively high presence of advice for 4 of the 5 types of advice in this phase. However, we could not observe any significance for these differences, which is why the sole presence of this outlier does not change the overall picture. 

\subsection*{Phase 5: Prevention}
\label{sec:prevention}

The final phase, which intends to prevent further malicious activity, consists of 9 different types of advice. In general, guidance in this phase is the least prominent ranging from 37\% to 43\%, as depicted in Figure~\ref{fig:prevention}. However, this is primarily due to one type of advice where services regularly inform users to make security checks ranging from only 4\% to 10\%. Again we observe the highest presence of this advice for popular websites in the Marshall Islands, Solomon Islands, and Vanuatu, the worst-ranked countries in Oceania. Nevertheless, we could not observe any significant differences here as it was already the case for the \textit{Service Restoration} phase. 

In contrast, helping users create a secure password is the most popular advice. At the same time, it also has the most substantial differences being present in 47\% of the websites popular in the worst-ranked countries in Oceania, whereas~79\% of those in the best-ranked countries in Asia and the worst-ranked in America provide this information. The $\chi^2$ test also confirmed this: advice for a secure password is significantly less present among popular websites in the worst-ranked countries in Africa and Oceania compared to both the worst- and the best-ranked countries in America, Asia, and Europe. 

\figOverall

\subsection*{Overall}
\label{sec:overall}

Now that all 5 phases have been analyzed separately, we also want to look at general trends. For this, Figure~\ref{fig:overview} presents the average across all 5 phases, i.e., all 31 codes. Here, the worst-ranked countries in Africa list the lowest with only~45\% while the best-ranked countries in Asia and Europe have the highest presence of advice with 51\%. When comparing the 2 groups within a continent, we see that the better-ranked countries usually also have a higher presence of advice, with the difference ranging from 2\% in Africa (47\% vs. 45\%) to~5\% in Asia (51\% vs. 46\%). Only in Oceania, the 3 lowest ranked countries have a slightly higher presence of advice,~48\% compared to 47\%. 

While the described observations are primarily marginal, there is a general tendency for countries that are part of the Global South to be more affected by the absence of account remediation advice. In our analysis, we also found multiple cases where this discrepancy becomes apparent, 2 of which were even significant: \textit{Billing/Finance Issues} in the first and \textit{Password Advice} in the fifth phase. We will discuss these findings and their implications in the subsequent sections. 

\section{Discussion}
\label{sec:discussion}
Our research analyzes account remediation for the most popular websites in Africa, America, Asia, Europe, and Oceania. In this section, we discuss the general outcome found on most websites regarding the procedures and protocols of account remediation, focusing on the differences between countries and continents.

\subsection{Situation of the Global South}
One aspect prevails throughout our analysis: the lack of advice on websites popular in the Global South is more distinct than in the Global North. For advice on financial or billing issues and creating a secure password, the difference is even significant when comparing the worst-ranked countries in Africa and Oceania to the other groups. Ultimately, this lack makes it harder for users to recover from a compromise which adds to the existing inequality between countries. On top of that, solving this problem is also more complex as countries in the Global South may have a different level of access to technology and internet connectivity than in the Global North, which could impact the organizations' ability to adjust their website. Furthermore, due to the prevalence of security advice in English, language barriers may also pose a problem. 

Increasing the presence of advice requires communicating the necessity for it, especially across continents, to improve the situation in countries other than, e.g., the U.S., Canada, and Europe. Unfortunately, communicating results and leading to a change can take time and effort. One of the most prominent examples of such a transition is the deprecation of password complexity rules. The British National Cyber Security Centre (NCSC) changed their guideline in 2016~\cite{ncsc-16-simplify}, the National Institute of Standards and Technology (NIST) in the USA followed in 2017~\cite{mcmillan-17-burr-was-wrong}, but still, only 13\% of websites adhered to them in 2022~\cite{lee-22-password-policies}. Moreover, most other countries still need federal agencies to communicate such changes, which makes conveying the importance of account remediation advice even more difficult.

\subsection{Account Deletion}
Nowadays, people have become aware of privacy and security problems, making them sometimes want to delete their accounts on websites they are no longer using. However, we found that websites make deleting accounts difficult. In our analysis, most websites do not provide enough information about the deletion mechanism; on average, 20\% do not provide information. Comparing websites that are popular in the best and worst-rated countries, we find that of the latter,~10\% contain fewer account deletion notices, regardless of the continent. Likewise, the websites do not mention what data will be deleted, whether it applies to the entire account, only personal data, or the user's activities. At the same time, some websites indicate not to delete any user data. Finally, there is a problem in determining the time of deletion because websites need to mention when the data will be removed from their databases. 

\subsection{Service Restoration}
Restoration protocols are among the essentials that must be available on websites in case the account has been hacked or deleted. Unfortunately, for most websites, we did not find any information or clear steps to restore an account in such a case. Usually, it falls back to a ``contact us if there are any problems.'' 
Furthermore, some websites only depend on sending a code to the email to verify the authentication and do not provide any alternatives, e.g., confirming a prompt on a linked device~\cite{google-20-phone-prompts}, providing a one-time or backup code~\cite{reese-19-comp-five-2fa}, or proving knowledge about the account such as the creation date or last login location~\cite{doerfler-19-login-challenges}.
This makes it only possible to restore the account if the email account is available.
 
\section{Implications and Recommendations}
\label{sec:implications}
Account remediation is not only a protocol that websites should follow; it is a complex way that brings in the socio-technical components to ensure data protection and prevent further data leakage for an account compromise. Additionally, the proper account remediation process enables users to protect their accounts as a preventive measure through secure behavior such as prevention of password rotation~\cite{parkin2015assessing,schechter-09-no-secret}. Our transcontinental analysis for the $158$ websites measures the account remediation for different aspects like changing passwords, reviewing past activities, enabling 2FA, and others. Such a large-scale study in this area is done outside our background research. With the study, we aim to add valuable feedback for evaluating the implementation of the account remediation steps, measuring the security of the websites across different countries, and assessing how websites follow the expert advice in various countries. Based on our evaluation, we provide several recommendations, which we outline below. 

The results we obtained confirm that most websites need more account remediation advice even across the globe, especially in the Global South. This highlights the need for more tailored solutions, information campaigns, and more focused research in the long term. To address the described issues, we suggest providing an easy-to-follow checklist for web developers to ensure they provide all account remediation steps. Similarly, a checklist for users could be an alternative to missing or addition to incomplete advice on websites. We discuss some of these recommendations in the following:

\subsection{Customer Support}
Most websites do not have a customer support and even if they do, they are primarily form-based without an immediate response. 
Nevertheless, it is crucial to provide customer support with immediate responses, not only to keep users satisfied but also to respond to security-related quickly and, thus, presumably time-critical issues, especially when users assign a high value to the account~\cite{wilson2001s}. 
If websites do not provide customer support, this is even more concerning because locked-out users cannot report an account compromise, take necessary steps to secure their account, or do a simple task such as changing the password. 

Hence, we encourage national and international policy-makers to mandate account remediation procedures for all websites that maintain user accounts.
This could be similar to breach notification obligations like the data breach notification laws in the US~\cite{ncsl-22-security-breach-laws} or GDPR in the EU~\cite{eu-16-gdpr}, which intend to give users the right to be made aware of data breaches which affect them.
However, it is essential to note that this requires substantial focus on improving the security workforce by the organization owning the website~\cite{newhouse2017national}. 
Therefore, it is crucial to focus on this workforce and provide proper customer support for the users to contribute to the account management. 

\subsection{Compromise Discovery Protocol}
We noticed room for improvement in compromise discovery and mitigation protocols, including flexibility in changing email and passwords or up-to-date password requirements. The central problem of the attacks is discovering the compromise concerning user data. For example, Acker et al. analyzed the top Alexa $100,000$ pages to identify login pages and measure their security. As a result, they found possibilities for password leaks to third parties and password eavesdropping on the network without the knowledge of the website owners~\cite{van2017measuring}. This is concerning as with such knowledge and a discovery mechanism in place; it is easier for the users to identify a compromise and act accordingly~\cite{diogenes2019cybersecurity,hosack2011businesses}. 

Therefore, websites should inform users about unusual activity, although this challenges them. 
Too many false positives, i.e., sending notifications for legit activity, can give misleading warnings to the users and create an illusion for the users' mental models to ignore the actual warning signs. Thus, websites need to be precise about what constitutes an ``unusual activity'' by giving concrete examples such as changed passwords or emails~\cite{markert-22-it-was-me, redmiles-19-should-worry}.
A second aspect of this is an interface to review account activities, such as logins and actions that alter an account's status (purchases, password or preference changes, settings, user information)~\cite{fernandes-22-device-activity-pages}
Through this interface, a user should be able to terminate any or all active sessions. Our results show that only a few websites currently provide this functionality: 33\% allow to sign out of individual sessions, 27\% allow to sign out everywhere, and numbers vary across countries and continents. Along with revealing account activity, there should also be an interface that allows users to delete undesired items and reverse changes made to their settings as suggested by prior work~\cite{bonneau-15-lies-and-account-recovery,huh2017m}.

\subsection{Robust Security Requirements}
As part of this study, we checked website password requirements and found that, on average, 53\% of them do not check for a proper password rotation~\cite{parkin2015assessing, schechter-09-no-secret}. 
This check is possible without any security infringement, as knowledge of the old password also needs to be double-checked. Otherwise, someone who does not know the old one could change the password but has access to the account, e.g., via a stolen session cookie or an existing session on a shared device.

Furthermore, 29\% do not advise users about strong passwords, and even more provide outdated requirements, which is in line with findings by Lee et al.~\cite{lee-22-password-policies}. For example, we have noted that banking websites such as JPMorgan Chase prevent password rotation. However, the same is not followed by the most common websites. 
Therefore, the usage of up-to-date password requirements as provided by the NIST SP800-63B~\cite{nist-17-sp800-63b} or the NCSC~\cite{ncsc-16-simplify} needs to be promoted even more. 
Strong passwords can be helpful to secure websites from many attacks like phishing, key logging, and brute-force attacks~\cite {florencio-07-strong-passwords}. Hence, passwords or, in general, authentication requirements should align with the expert advice like usage of 2FA~\cite{das2020non,das2019towards,das2020don,patrick2022understanding,rader2015identifying,zurko2005user}. Still, even those websites with 2FA enabled primarily activated it once during account creation and often did not have any mechanism to protect user data after a device change.

\subsection{Account Deletion, Recovery, and Restoration}
A proper data deletion protocol is necessary to review which data was removed from user accounts~\cite{reardon2013sok}. We found that many websites need more information on what data will be removed when the users delete their accounts. So, we recommend that websites have information about deleting users' data, how users can delete their accounts, and how long it would take to delete the entire accounts from the database. 

Regarding recovering and restoring the service for any account compromise, we found that email is the most used communication channel. 
Note that phone calls are susceptible to phishing attacks~\cite{banu2013comprehensive}. 
Thus, email can also be considered the more secure communication channel~\cite{innocenti2021you} unless dealing with financial institutions or when critical information is exchanged where a high level of security is necessary, e.g., in-person or physically mailing to the user address. 
In addition, we considered security verification checks that can be a second step to confirm the user's identity to restore an account. 
Again, we found that email is the most used channel for this type of check. 
Hence, we also want to stress the importance of assuring the security of this connection by using state-of-the-art configurations, including DNSSEC, SPF, DKIM, DMARC, DANE, and STARTTLS~\cite{shen-21-email-spoofing}.

\subsection{Proper Documentation and Better Guidance}
In general, we noted a lack of information and documentation provided by the websites, which became even more evident for the Global South. 
Still, even when the instructions for account remediation were available, obtaining specific information, e.g., about how to delete the account or reset the password, took work. 
Additionally, we discovered significant heterogeneity in the preventative documentation and guidance offered by these websites, indicating that many need more prevention guidance, which can harm users. 
Prior research has shown that such guidance and proper documentation are invaluable for user data protection~\cite{redmiles-16-advice-digital-security,redmiles2020comprehensive}; however, several websites disregard that. 
Various websites need to provide some of the features of their web application counterparts. 
Though this was out of the scope of our current research study, we plan to explore this disparity as a future extension of this work.
\section{Limitations and Future Work}
\label{sec:limit}
Account remediation is an essential aspect of assessing the protocols used in the websites to protect the users' data and accounts. However, despite our best efforts, we needed help analyzing the websites for the following reasons. First, we analyzed websites from different continents. Consequently, the first language for many of these websites is not English, and some are not offered in English at all. Therefore, to analyze non-English pages, we used translation services, e.g., the built-in feature in Google Chrome. Still, not all of these translations were perfect, which may have influenced the analysis. Secondly, some websites do not allow account sign-up, such as online news, sports news, online platforms, games, and streaming websites. As a result, we limited ourselves to websites that allow account sign-up.

Moreover, we found that some websites need to allow us to access them from our location. We tried to circumvent any location-based blocks by using a VPN, but some websites remained unavailable. Finally, to facilitate replication, we did not analyze websites that required a national ID, credit card number, phone number, or company name to register. As such websites occurred for each country, we expect any potential differences due to this constraint to be negligible.

Based on our analysis of the top 50 websites for each country, we already observed significant differences for a specific account for remediation steps when comparing countries and continents. However, we assume the differences to be even more noticeable when including more and less popular websites. A broader analysis covering more countries and websites is needed to confirm this assumption.
Finally, we plan to do a time-based analysis to provide a detailed evaluation of the account remediation protocols. Through such an analysis, we can provide both the perspective of account remediation protocols offered by the website and how seamlessly it is implemented in everyday usage.

\section{Conclusion}
\label{sec:conclusion}
\textit{Account Remediation} is the protective mechanism that ensures the account access is restricted and the data leakage is prevented by following the protocols defined. Thus, the account remediation process is technically complex, requiring several procedures to ensure legitimate access is not restricted and illegitimate access and data leakage are prevented. Thus, in most cases, this procedure is left to the end-user knowledge and proactive behavior. However, many websites either do not create the technical feasibility to conduct a proper account remediation protocol or provide adequate documentation for users to be aware of the procedure they need to do to secure their accounts. 

Therefore, our goal for this study was to conduct an intercontinental analysis of the account remediation protocols of the top $50$ websites from 6 countries on 5 continents: Africa, America, Asia, Europe, and Oceania. Of the websites, 3 of the 6 were the best performing, and the remaining 3 were the worst countries based on the cybersecurity index from GCI. We analyzed the websites based on the 5 phases of account remediation: compromise discovery, account recovery, limit access, service restoration, and prevention. As part of this, we note features such as the flexibility of changing passwords, identifying unwanted activities, deleting accounts, and enabling 2FA. Our analysis showed that websites across the globe do not follow expert advice as part of the account remediation schema. 
For example, initially identifying a compromise is difficult, websites require their users to be able to log in to take countermeasures, and password advice needs to be updated or included. 
When comparing countries and continents, this need for adequate information is even more prevalent in the Global South, showing a technological disparity compared with the Global North. This highlights the need for easy-to-follow advice that reaches users and website providers alike, which we intend to prepare, test and provide as part of future work.

\section*{Acknowledgment}
We would like to thank the Inclusive Security and Privacy-focused Innovative Research in Information Technology (InSPIRIT) Laboratory at the University of Denver. We would also like to thank Hebah Alafari, Ahmed Khormi, Shatha Surayyi, and Alyssa Zinn for their initial contribution in this research. Any opinions, findings, conclusions, or recommendations expressed in this material are solely those of the authors. They do not necessarily reflect the views of any organization with which the authors may be affiliated.

\newpage
\balance
\bibliographystyle{IEEEtranS}
\bibliography{bibfile}

\onecolumn
\appendix

\label{sec:additionaltables}
\tableAllWebsitesPartOne{}
\tableAllWebsitesPartTwo{}

\end{document}